\documentclass[10pt,twocolumn,twoside]{IEEEtran}

\usepackage{cite}
\usepackage[dvips]{graphicx}
\usepackage{amsmath}
\usepackage{amssymb} 
\usepackage{algorithm}
\usepackage{algcompatible}
\usepackage[caption=false,font=footnotesize]{subfig}
\usepackage{bm}
\usepackage{forest}
\usepackage{amsthm}
\usepackage{circuitikz}

\theoremstyle{plain}
\newtheorem{theorem}{Theorem}

\newtheorem{corollary}{Corollary}

\DeclareMathOperator*{\argmin}{arg\,min}

\DeclareMathOperator{\blockdiag}{blockdiag}

\DeclareMathOperator{\card}{card}
\DeclareMathOperator{\ve}{\mathrm{vec}}
\renewcommand{\Re}{\operatorname{Re}}
\renewcommand{\Im}{\operatorname{Im}}

\newcommand{\RR}{\mathbb{R}}
\newcommand{\Co}{\mathbb{C}}

\newcommand{\ii}{\mathbf{i}}
\newcommand{\vv}{\mathbf{v}}
\newcommand{\ff}{\mathbf{f}}
\newcommand{\ov}{\overline{\mathbf{v}}}
\newcommand{\xx}{\mathbf{x}}

\newcommand{\FF}{\mathbf{F}}
\newcommand{\oF}{\overline{\mathbf{F}}}
\newcommand{\YY}{\mathbf{Y}}
\newcommand{\II}{\mathbf{I}}
\newcommand{\VV}{\mathbf{V}}
\newcommand{\ZZ}{\mathbf{Z}}

\newcommand{\QQ}{\mathbf{Q}}
\newcommand{\Tr}{^{\top}}

\newcommand{\uZ}{\mathbf{\underline{0}}}
\newcommand{\ze}{\mathbf{0}}
\newcommand{\Id}{\bm{\mathcal{I}}}

\renewcommand{\SS}{\mathcal{S}}
\newcommand{\PP}{\mathcal{P}}

\newcommand{\NN}{\mathcal{N}}

\newcommand{\Vc}{\bm{\mathcal{V}}}
\newcommand{\cc}{\mathbf{c}}
\newcommand{\dd}{\mathbf{d}}

\newcommand{\LL}{\mathcal{L}}
\newcommand{\TT}{\mathcal{T}}

\newcommand{\zz}{\mathbf{z}}
\newcommand{\yy}{\mathbf{y}}

\begin{document}

\title{Estimating the Frequency Coupling Matrix from Network Measurements}

\author{Antoine~Lesage-Landry,~\IEEEmembership{Student Member,~IEEE}, Siyu Chen,~\IEEEmembership{Student Member,~IEEE}, and~Joshua~A.~Taylor,~\IEEEmembership{Member,~IEEE}%
\thanks{This work was funded by the Fonds de recherche du Qu\'ebec -- Nature et technologies, the Ontario Ministry of Research, Innovation and Science and the Natural Sciences and Engineering Research Council of Canada.}%
\thanks{A. Lesage-Landry, S. Chen and J.A. Taylor are with The Edward S. Rogers Sr. Department of Electrical \& Computer Engineering, University of Toronto, Toronto, Ontario, Canada, M5S 3G4.
        \{\texttt{alandry@ece.,siyuuu.chen@mail.} \texttt{,josh.taylor@}\}\texttt{utoronto.ca}}
}

\maketitle

\begin{abstract}
Power converters are present in increasing numbers in the electric power grid. They are a major source of harmonic currents and voltages, which can reduce power quality and trip protection devices. The frequency coupling matrix (FCM) is a general technique for modeling converter harmonics. It can be obtained through experimental characterization or, given a converter's internal parameters, direct calculation. In this paper, we estimate FCMs from network measurements. We give a novel harmonic reduction theorem for computing equivalent, virtual FCMs for unobservable portions of a network. We estimate FCMs and harmonic line admittances from PMU measurements, and give an efficient online version for the FCM problem. We test our approaches on a numerical example, and show that the estimation error is low under noisy observations.
\end{abstract}


\IEEEpeerreviewmaketitle

\section{Introduction}
\IEEEPARstart{P}{ower-electronic} converters are required to integrate renewables like photovoltaics and wind turbines into the grid~\cite{arrillaga2004power,carrasco2006power,masters2013renewable}. Converters approximate sinusoidal signals with switched signals, which introduces harmonics into the power system~\cite{arrillaga2004power,de2006harmonics}. Harmonics have a number of undesirable consequences, including reduced power quality, increased losses, and vibration in mechanical equipment~\cite{henderson1994harmonics,arrillaga2004power,gray2012time,lian2012steady}.

Harmonics are difficult to model accurately because they are the result of numerous non-idealities. The frequency coupling matrix (FCM) is a powerful technique for modeling converter harmonics because it can be computed directly from converter parameters~\cite{rajagopal1993harmonic,fauri1997harmonic,sun2007harmonically,lehn2007frequency}, or empirically from laboratory measurements~\cite{yahyaie2014application,zong2016new,yahyaie2016using}. In this paper, we extend the empirical approach to the network setting.

In this work, we formulate least squares problems for estimating the FCM and harmonic line admittances from phasor measurement units (PMUs). We give a novel network reduction theorem for representing unobservable portions of the network with an equivalent virtual FCM. This information can be used in several applications such as power flow analysis and for the placement of fault protection devices~\cite{saadeh2016estimation}. It can also facilitate the control of harmonic injected in the network, for example, by incorporating harmonics in optimal power flow~\cite{tian2017harmonic}. 

We now review the relevant literature. A number of authors have computed the FCM using the converter's modulation characteristics~\cite{larsen1989low,jalali1991harmonic,rajagopal1993harmonic,saniter2003modelling,hu1992harmonic}. These approaches have a limited reach for network estimation as they cannot be used without exact knowledge of the converters' internal parameters. In~\cite{lehn2007frequency}, an analytical calculation of the FCM is given for converters in steady state. References~\cite{yahyaie2014application,zong2016new,yahyaie2016using} describe a method for obtaining the FCM of an individual converter experimentally and with no knowledge of internal parameters. This experimental approach is accurate, but requires offline measurements prior to the installation of the power converter. It also falls short when the FCM changes through time due to modifications in its operational parameters like the switching times or input dc current. 

Several approaches have been proposed to compute the line admittance matrix of a network. In~\cite{bazrafshan2018comprehensive}, the admittance matrix is determined experimentally from known subnetwork models such as transmission lines, transformers and step-voltage regulators. In~\cite{ardakanian2017identification}, the fundamental frequency admittance matrix is estimated using least squares. They also consider topology identification. We extend the least squares estimation portion of~\cite{ardakanian2017identification} to include harmonic frequencies as well.

In this paper, we estimate the FCM and line admittances from PMU measurements. Signal processing tools such as the fast Fourier transform can be used to decompose the voltage and current measurements into fundamental and harmonic phasors~\cite{carta2009pmu,jain2017fast,melo2017harmonic}. The maximum harmonic order $K$ depends on the sampling rate of the PMU. Given the standard 48 samples per cycle PMU, harmonic phasors up to $K=24$ can be computed for a fundamental frequency of $60$ Hz. The sampling rate of more recent PMUs is as high as 128 samples per cycle~\cite{phadke2008synchronized}, enabling harmonic analysis up to $K=64$. According to Standard C37.118-2005, a PMU should be able to transmit at a rate between 10 Hz and half its nominal frequency~\cite{martin2008exploring}. This would provide an adequate flow of measurements to estimate the FCM in real-time.

Our contributions are the following:
\begin{itemize}
  \item We give a network reduction theorem, which enables us to represent unobservable portions of the network with an equivalent, virtual FCM;
  \item We formulate a least squares problem for estimating harmonic line admittances from network measurements;
  \item We formulate a least squares problem for estimating FCMs from network measurements;
  \item We give an efficient online algorithm for cases in which the FCM is time-varying;
  \item We validate the reduction theorem and solve the least squares estimation problems in a numerical example.
\end{itemize}

\section{Notation \& Background}

\subsection{Harmonic network}
We consider a three-phase network and model harmonic frequencies up to the $K^\text{th}$ order. We denote the set of nodes of the network by $\NN = \{1, 2, \ldots, N\}$ and the set of transmission lines by $\mathcal{M} \subseteq \mathcal{N\times{N}}$. 

Let $\ii^k_{n,t} \in \Co$ and $\vv^k_{n,t} \in \Co$ be the harmonic current and voltage for the $k^\text{th}$ harmonic at node $n \in \NN$ and time $t$. For a node $n$ with a power converter, $\ii^k_{n,t}$ and $\vv^k_{n,t}$ represent the current and voltage on the grid side of the converter. Let $\ii_{n,t,\text{dc}}^k \in \Co$ and $\vv_{n,t,\text{dc}}^k \in \Co$ be the $k^\text{th}$ harmonic current and voltage on the dc side of the converter. Let $\zz_{n,m}^k \in \Co$ and $\yy^k_{n,m} \in \Co$ denote the impedance and admittance at harmonic frequency $k$ between node $n$ and $m$ for $(n,m) \in \mathcal{M}$. We denote individual phases by appending $(a)$, $(b)$ or $(c)$ to a variable or parameter.

\subsection{Line Admittance}
At frequency $k$, phase $a$ and time $t$, we have
\[
\ii_{\text{bus},t}^k(a) = \YY_\text{line}^k(a) \vv_{\text{bus},t}^k(a).
\]
Combining the three phases we obtain
\[
\ii_{\text{bus},t}^k = \YY_\text{line}^k \vv_{\text{bus},t}^k,
\]
where $\YY_\text{line}^k= \blockdiag\left(\YY_\text{line}^k(p), p = a,b,c\right)$, $\ii_{\text{bus},t}^k = \begin{pmatrix}
\ii_{\text{bus},t}^k(a) & \ii_{\text{bus},t}^k(b) & \ii_{\text{bus},t}^k(c)
\end{pmatrix}\Tr$ and similarly for $\vv_{\text{bus},t}^k$.

Let $u = 3N(K+1)$. Define the block diagonal harmonic admittance matrix $\YY_\text{H} \in \Co^{u \times u}$ as:
\[ 
\YY_\text{H}= \blockdiag\left( \YY_\text{line}^0, \YY_\text{line}^1, \YY_\text{line}^2, \ldots, \YY_\text{line}^K \right). 
\] 
We can then write
\begin{equation}
\begin{pmatrix}
\ii_{\text{bus},t}^0 \\
\ii_{\text{bus},t}^1 \\
\ii_{\text{bus},t}^2 \\
\vdots \\
\ii_{\text{bus},t}^K\\
\end{pmatrix}=\YY_\text{H}
\begin{pmatrix}
\vv_{\text{bus},t}^0 \\
\vv_{\text{bus},t}^1 \\
\vv_{\text{bus},t}^2 \\
\vdots\\
\vv_{\text{bus},t}^K \\
\end{pmatrix}. \label{eq:Y_H}
\end{equation}
We write~\eqref{eq:Y_H} in short form as:
\[
\ii_{\text{bus},t} = \YY_\text{H} \vv_{\text{bus},t}.
\]
Note that the off-diagonal terms in $\YY_\text{H}$ are neglected due to the weak linkage between different harmonics in the lines.

\subsection{Frequency coupling matrix}
The frequency coupling matrix models the harmonics generated by power converters~\cite{rajagopal1993harmonic,larsen1989low,lehn2007frequency,sun2007harmonically}. We use the formulation of~\cite{lehn2007frequency}. At a power converter, the FCM, $\mathbf{{\widetilde{F}}}$ relates the harmonic currents and voltages as
\begin{equation}
\begin{pmatrix}
\ii^{H}(a) \\ \ii^{H}(b) \\ \ii^{H}(c)\\ \vv_\text{dc}^{H}
\end{pmatrix} = \mathbf{{\widetilde{F}}}
\begin{pmatrix}
\vv^{H}(a) \\ \vv^{H}(b)\\ \vv^{H}(c) \\ \ii_\text{dc}^{H}
\end{pmatrix} , \label{eq:fcm_complex}
\end{equation}
where 
\begin{align*}
\ii^H =& \left(
\ii^{-K} \; \ii^{-K+1} \; \ldots \; \ii^{-1} \; \ii^{0} \; \ii^{1} \; \ldots \; \ii^{K-1} \; \ii^{K}
\right)\Tr,\\
\vv^H =& \left(
\vv^{-K} \; \vv^{-K+1} \; \ldots \; \vv^{-1} \; \vv^{0} \; \vv^{1} \; \ldots \; \vv^{K-1} \; \vv^{K}
\right)\Tr.
\end{align*}
Note that, due to the definition of phasors, $\ii^{-k}$ and $\vv^{-k}$ are the complex conjugates of $\ii^k$ and $\vv^k$.

Our objective is to estimate the FCM in order to model the harmonics injected in the network by the converter. Hence, we drop the dc voltage in the left-hand side of~\eqref{eq:fcm_complex}, which is typically very tightly regulated. Furthermore, we neglect all harmonic currents on the dc side of the converter because they are typically very small in magnitude~\cite{lehn2007frequency,zong2016new,tian2017harmonic}. These assumptions lead to
\begin{equation}
\begin{pmatrix}
\ii^{H}(a) \\ \ii^{H}(b) \\ \ii^{H}(c)
\end{pmatrix} = \mathbf{\hat{F}}
\begin{pmatrix}
\vv^{H}(a) \\ \vv^{H}(b)\\ \vv^{H}(c) \\ \ii_\text{dc}^{0}
\end{pmatrix} , \label{eq:fcm_pre_final}
\end{equation}
where $\mathbf{\hat{F}}$ is the matrix $\mathbf{{\widetilde{F}}}$ with the appropriate rows and columns removed. Without loss of generality, we rewrite~\eqref{eq:fcm_pre_final} as a real-valued equation using the transformation detailed in~\cite[Appendix D]{lehn2007frequency}, 
leading to
\begin{equation}
\begin{pmatrix}
\Re\left(\ii^{0}(a) \right) \\ \Im\left(\ii^{0}(a) \right) \\ \Re\left(\ii^{1}(a) \right) \\ \Im\left(\ii^{1}(a) \right) \\ \vdots \\ \Re\left(\ii^{K}(a) \right) \\ \Im\left(\ii^{K}(a) \right)  \\ \Re\left(\ii^{0}(b) \right) \\ \Im\left(\ii^{0}(b) \right) \\ \vdots \\ \Re\left(\ii^{K}(c) \right) \\ \Im\left(\ii^{K}(c) \right)
\end{pmatrix} = \FF
\begin{pmatrix}
\Re\left(\vv^{0}(a) \right) \\ \Im\left(\vv^{0}(a) \right) \\ \Re\left(\vv^{1}(a) \right) \\ \Im\left(\vv^{1}(a) \right) \\ \vdots \\ \Re\left(\vv^{K}(a) \right) \\ \Im\left(\vv^{K}(a)\right)  \\ \Re\left(\vv^{0}(b) \right) \\ \Im\left(\vv^{0}(b)\right) \\ \vdots \\ \Re\left(\vv^{K}(c) \right) \\ \Im\left(\vv^{K}(c)\right) \\ i_\text{dc}^{0}
\end{pmatrix}. \label{eq:fcm_approx}
\end{equation}
We write~\eqref{eq:fcm_approx} in condensed form by
\begin{equation}
\ii = \FF \vv, \label{eq:fcm_final}
\end{equation}
where $\ii$, $\vv$ and $\FF$ are real-valued. Observe that the last entry of the vector $\vv$ is $i_\text{dc}^0$, a real-valued current. 

Also, note that $\vv^k_\text{dc}$ for $k = 0, 1, \ldots , K$ can be included in~\eqref{eq:fcm_final} if this quantity is of interest. In this case, all of the following sections except~\ref{sec:network_red} apply.

\subsection{Observability}

A network is observable if the available measurements allow the computation of a unique voltage phasor at every node~\cite{abur2004power,krumpholz1980power}. An observable island is a portion of the network that is fully observable, and a node is observable if its voltage phasor can be uniquely estimated. The definitions extend straightforwardly to our setting due to the linearity of the lines. See~\cite[Chapter 4]{abur2004power} for methods to determine the observability of a network.

\section{Network reduction}
\label{sec:network_red}

We now present a result that allows us to model the harmonics of an arbitrary subtree with a single, virtual FCM. This is useful when portions of a network are unobservable, e.g., due to lack of PMUs. Let $\FF_\SS$ be the virtual FCM for a subtree $\SS$. If no converters are part of the tree, the dc current is zero and the FCM $\FF_\SS$ is the equivalent load admittance matrix of the tree as no coupling between frequencies occur. The results of this section enable us to compute the $\FF_\SS$ using only current and voltage measurements at the root of $\SS$.

Before stating our results, we give a few definitions. For any voltage and dc current vector $\vv$, we define $\ov \in \RR^{6(K+1)}$ such that
\begin{equation}
\vv =
\begin{pmatrix}
\ov \\ i^0_\mathrm{dc}
\end{pmatrix}.
\label{eq:v}
\end{equation}
Similarly, we define $\oF \in \RR^{6(K+1) \times 6(K+1)}$ and $\ff \in \RR^{6(K+1)}$ such that
\begin{equation}
\FF =
\begin{pmatrix}
\oF & \ff
\end{pmatrix}.
\label{eq:F}
\end{equation}
Let $\mathbf{\hat{Z}}_{n,m} \in \Co^{3(2K+1) \times 3(2K+1)}$ be a diagonal matrix of the impedances between two adjacent nodes $n,m \in \NN$ for all harmonic frequencies $k=-K, -K +1, \ldots , -1, 0,1, \ldots, K$ and phases $a$, $b$ and $c$. Let $\ZZ_{n,m}  \in \RR^{6(K+1) \times 6(K+1)}$ be the real-valued impedance matrix obtained by applying~\cite[Appendix D]{lehn2007frequency} to $\mathbf{\hat{Z}}_{n,m}$.

\begin{theorem}
Let $\SS$ be a tree of maximum depth one. Let $s$ be the root node and $\SS^\ast = \SS \backslash \{s\}$ be the leaf nodes. Let $\PP$ be the set of converters connected to $s$. Suppose there is a power converter at each node $n \in  \SS^\ast$ with  FCM $\FF_n$. Assume $\mathbf{M}_{s,n} = \ZZ_{s,n} \oF_n + \mathbf{I} $ is invertible for all $n \in \SS^\ast$. Then there exists an FCM for $\SS$, $\FF_\SS$, such that $\ii_s = \FF_\SS \vv_s$ for any feasible voltage and current at $s$. The FCM $\FF_\SS$ is fully determined by the line impedances, FCMs, and dc currents in the subtree $\SS$.
\label{thm:net_reduc}
\end{theorem}

\begin{IEEEproof}
Suppose that $\ii_s$ and $\vv_s$ are the voltage and current at the root node, $s$. We show by construction that there exists a unique matrix $\FF_\SS$ such that $\ii_s = \FF_\SS \vv_s$.

Evaluating Kirchhoff's current law at node $s$ gives
\begin{equation}
\ii_s = \sum_{p \in \PP} \ii_p + \sum_{n \in \SS^\ast} \ii_n.
\label{eq:sum_currents}
\end{equation}
For all $p \in \PP$ and $ n\in \SS^\ast$, substituting~\eqref{eq:fcm_final} in~\eqref{eq:sum_currents} gives
\begin{equation}
\ii_s = \sum_{p \in \PP} \FF_p \vv_p + \sum_{n \in \SS^\ast} \FF_n \vv_n.
\label{eq:all_currents}
\end{equation}
We now use the definitions~\eqref{eq:v} and~\eqref{eq:F} in~\eqref{eq:all_currents}. The voltage $\ov_p=\ov_s$ for all $p \in \PP$ because they are at the same bus. We re-express the first sum of the right-hand side as
\begin{align}
\ii_s &= \sum_{p \in \PP} \begin{pmatrix}
\oF_p & \ff_p
\end{pmatrix} \begin{pmatrix}
\ov_s \\ \ii^0_{\mathrm{dc},p}
\end{pmatrix} + \sum_{n \in \SS^\ast} \FF_n \vv_n \nonumber \\
&= \underbrace{\sum_{p \in \PP} \oF_p}_{\oF_\PP} \ \ov_s + \underbrace{\sum_{p \in \PP} \ff_p \ii^0_{\mathrm{dc},p}}_{\ff_\PP}+ \sum_{n \in \SS^\ast} \FF_n \vv_n  \nonumber\\
&= \oF_\PP \ov_s + \ff_\PP + \sum_{n \in \SS^\ast} \FF_n \vv_n.
\label{eq:inter_current}
\end{align}
We use Ohm's Law to relate the voltage at $s$ with the voltage at the leaf nodes. For $n \in \SS^\ast$, we have
\[
\ov_s - \ov_n = \ZZ_{s,n} \ii_n.
\]
Using~\eqref{eq:fcm_final} for all $n \in \SS^\ast$ leads to
\begin{align*}
\ov_s  &= \ZZ_{s,n} \FF_n \vv_n + \ov_n \\
&= \ZZ_{s,n} \FF_n \begin{pmatrix}
\ov_n \\ i^0_{n,\mathrm{dc}}
\end{pmatrix} + \ov_n \\
&= \ZZ_{s,n} \left( \oF_n \ov_n + \ff_n i^0_{n,\mathrm{dc}} \right) + \ov_n.
\end{align*}
Rearranging the terms, we have
\[
\ov_s = \left( \ZZ_{s,n} \oF_n + \mathbf{I} \right) \ov_n+ \ZZ_{s,n} \ff_n i^0_{n,\mathrm{dc}}.
\]
Now recall that $\mathbf{M}_{s,n} = \ZZ_{s,n} \oF_n + \mathbf{I}$. By assumption, $\mathbf{M}_{s,n}$ is invertible. Solving for $\ov_n$, we have
\begin{align}
\ov_n &= \mathbf{M}_{s,n}^{-1} \left(\ov_s - \ZZ_{s,n} \ff_n i^0_{n,\mathrm{dc}}\right). \label{eq:ov_n_func}
\end{align}
Let $\bm{\ell}$ be the last term of the right-hand side of~\eqref{eq:inter_current}. We first re-express $\bm{\ell}$ in term of $\ov_n$, and then substitute it into~\eqref{eq:ov_n_func}. We have
\begin{align}
\bm{\ell} &= \sum_{n \in \SS^\ast} \FF_n \begin{pmatrix}
\ov_n \\ i^0_{n,\mathrm{dc}}
\end{pmatrix} \nonumber\\
&= \sum_{n \in \SS^\ast} \FF_n \begin{pmatrix}
\mathbf{M}_{s,n}^{-1} \left(\ov_s - \ZZ_{s,n} \ff_n i^0_{\mathrm{dc},n}\right) \\
 \ii^0_{n,\mathrm{dc}}
\end{pmatrix}\nonumber\\
&= \sum_{n \in \SS^\ast} \begin{pmatrix}
\oF_n & \ff_n
\end{pmatrix} \begin{pmatrix}
\mathbf{M}_{s,n}^{-1} \ov_s -  \mathbf{M}_{s,n}^{-1} \ZZ_{s,n} \ff_n i^0_{n,\mathrm{dc}} \\
 i^0_{n,\mathrm{dc}}
\end{pmatrix} \nonumber\\
&= \sum_{n \in \SS^\ast}  \oF_n \mathbf{M}_{s,n}^{-1} \ov_s - \oF_n \mathbf{M}_{s,n}^{-1} \ZZ_{s,n} \ff_n i^0_{n,\mathrm{dc}} + \ff_n i^0_{n,\mathrm{dc}} \vphantom{\oF_n^{-1}}  \nonumber\\
&= \sum_{n \in \SS^\ast}  \left(\ff_n - \oF_n \mathbf{M}_{s,n}^{-1} \ZZ_{s,n} \ff_n\right)i^0_{n,\mathrm{dc}} + \oF_n \mathbf{M}_{s,n}^{-1} \ov_s.
\label{eq:last}
\end{align}
Define
\begin{align*}
\hat{\ff}_\SS &= \sum_{n \in \SS^\ast}  \left(\ff_n - \oF_n \mathbf{M}_{s,n}^{-1} \ZZ_{s,n} \ff_n\right) i^0_{n,\mathrm{dc}}, \\
\hat{\oF}_\SS &= \sum_{n \in \SS^\ast} \oF_n \mathbf{M}_{s,n}^{-1}.
\end{align*}
We can then write~\eqref{eq:last} as
\[
\bm{\ell}= \hat{\oF}_\SS \ov_s + \hat{\ff}_\SS.
\] 
Substituting $\bm{\ell}$ into~\eqref{eq:inter_current}, we obtain
\begin{align*}
\ii_s &= \oF_\PP \ov_s + \ff_\PP +\hat{\oF}_\SS \ov_s + \hat{\ff}_\SS \\
\ii_s &= \underbrace{\begin{pmatrix}
\oF_\PP + \hat{\oF}_\SS & \ff_\PP + \hat{\ff}_\SS
\end{pmatrix}}_{\FF_\SS}
\begin{pmatrix}
\ov_s \\
1
\end{pmatrix} \\
&= \FF_\SS \vv_s.
\end{align*}
where $\vv_s = \left( \ov_s \; 1 \right)\Tr$. This establishes the existence of the matrix $\FF_\SS$. Observe that because we assume that $\mathbf{M}_{s,n}$ is full rank for all $n \in \SS^\ast$, $\FF_\SS$ is uniquely determined by the above construction. 
\end{IEEEproof}

Note that if several power converters are connected to $n \in \SS^\ast$, then Theorem~\ref{thm:net_reduc} can be first applied with $n$ as the root node and $\SS^\ast = \emptyset$ to obtain an equivalent FCM for a single, virtual converter. Also note that there is no physical quantity for the dc current of the virtual FCM. For this reason, we set the resulting mathematical quantity, i.e., the last element of $\vv_\SS$, to one. The next result generalizes Theorem~\ref{thm:net_reduc} for trees of depth greater than one.

\begin{corollary}[General tree]
Let $\TT$ be a tree with power converters at each node $n \in \TT$. Assuming that $\mathbf{M}_{m,n}$ is invertible for any pair of adjacent nodes $m,n \in \TT$, then there exists an equivalent, virtual FCM $\FF_\TT$ for the tree $\TT$ and it is unique.
\label{cor:tree_reduc}
\end{corollary}

\begin{IEEEproof}
We prove this corollary by iteratively applying Theorem~\ref{thm:net_reduc}. Let $\mathcal{T}_r$ be the reduced tree after Steps~\ref{en:step_leaf} and~\ref{en:step_subtree} have been applied $r$ times and let $\mathcal{L}_r \subseteq \mathcal{T}_r$ be the set of leaf nodes of $\mathcal{T}_r$. Let $\PP_n$ be the set of power converters connected to node $n$ and $\SS_n$ be the subtree with root node $n$. We obtain $\FF_{\mathcal{T}}$ via the following steps.
\begin{enumerate}
  \item For each $l \in \LL_r$ such that $\mathrm{card}\left(\PP_l \right) > 1$, apply Theorem~\ref{thm:net_reduc} with $s=\ell$ and $\SS_n^\ast =\emptyset$. This ensures that all leaf nodes are associated with a single FCM.
  \label{en:step_leaf}
  \item For all subtrees $\SS_n$ of depth 1 
  comprised only of nodes $l \in \LL_r$ and a parent node $n$, apply Theorem~\ref{thm:net_reduc}. 
  \label{en:step_subtree}
  \item Update $\TT_r$ and $\LL_r$. 
  \label{en:update}
\end{enumerate}
Repeating Steps~\ref{en:step_leaf}-\ref{en:update} will eventually reduce the network to a single node described by a single, virtual FCM, which we denote $\FF_\TT$.\hfill \IEEEQEDhere
\end{IEEEproof}

Note that the FCM of a node without a converter is equivalent to the node's load admittance matrix, appropriately formatted to match our notation.

\begin{figure}[!t]
\centering
\subfloat[Full tree with subtree $\mathcal{S}$]{
\centering
\begin{forest}
for tree = {
        l=1cm,
        s sep=1.35cm,
        fill=lightgray
        }
 [,circle,draw,fill=black,label={$a$}
 [1,circle,draw][2,circle,draw,name=top [3,circle,draw,name=bottomleft][4,circle,draw,name=bottomright]]
 ]
\node[draw=gray,dashed,thick,rounded corners=0.25cm,fit=(top.north east)(bottomleft)(bottomright),label={$\mathcal{S}$}] {};
\end{forest}%
\label{fig:full}} \quad
\subfloat[Equivalent reduced tree]{
\centering
\begin{forest}
for tree = {
        l=1cm,
        s sep=1.35cm,
        fill=lightgray
        }
 [,circle,draw,fill=black,label={$a$}
 [1,circle,draw][$\mathcal{S}$,circle,draw,fill=white,name=top 
 [,circle,fill=white,no edge,name=bottomleft][,circle,fill=white,no edge,name=bottomright]]
 ]
\node[draw=white,dashed,thick,rounded corners=0.25cm,fit={(top) (bottomleft) (bottomright)}] {};
\end{forest}%
\label{fig:reduced}}
\caption{Network reduction (\emph{gray}: nodes with power converters, \emph{white}: node with an equivalent FCM, \emph{black}: regular nodes)}
\label{fig:fi_reg}
\end{figure}
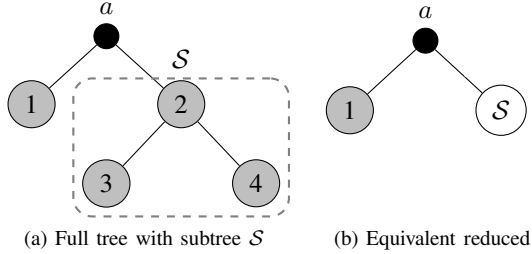

A visual representation of the network reduction process is shown in Figure~\ref{fig:fi_reg}. Given the FCMs and impedances of all components in subtree $\SS$ in Figure~\ref{fig:full}, Theorem~\ref{thm:net_reduc} enables us to compute an equivalent, virtual FCM, $\FF_\SS$, shown in Figure~\ref{fig:reduced}. Theorem~\ref{thm:net_reduc} could then be applied a second time on the reduced tree to calculate the equivalent FCM for the entire tree. 

This result also enables us to estimate unobservable portions of a network. Suppose that the only measurements available for subtree $\SS$ in Figure~\ref{fig:full} are the voltage and current at node $2$. One can then formulate an estimation problem for the virtual FCM, $\FF_\SS$. If there are enough measurements to guarantee a unique solution to the estimation problem, then Theorem~\ref{thm:net_reduc} guarantees that the result will be the unique, physically correct virtual FCM for subtree $\SS$.

\section{Line admittance estimation}

In this section, we formulate a least squares estimation problem for the line admittances. We assume that we have access to direct measurement at all nodes of the network~\cite{ardakanian2017identification} and that we have access to sufficient measurement data from PMUs. The harmonic admittance matrix estimation problem takes the following form:
\begin{equation}
\begin{aligned}
& \min_{\YY_\text{H} \in \RR^{u \times u}} & & \sum_{t=1}^T \left\|\ii_{\text{bus},t} - \YY_\text{H} \vv_{\text{bus},t} \right\|_F^2. \\
\end{aligned} \label{eq:ad_esti_pre}
\end{equation}
Let $\II_\text{network} \in \Co^{u \times T}$ and $\VV_\text{network} \in \Co^{u \times T}$ be the voltage and current measurement for all nodes, harmonics, and times. The estimation problem~\eqref{eq:ad_esti_pre} can be rewritten equivalently as
\begin{equation}
\begin{aligned}
\min_{\YY_\text{H} \in \RR^{u \times u}} \left\|\II_\text{network} - \YY_\text{H} \VV_\text{network} \right\|_F^2,
\label{eq:min_admit}
\end{aligned}
\end{equation}
where $\| \cdot \|_F$ denotes the Frobenius norm. The admittance matrix is highly sparse due to the topology of the power network. It is also by definition a symmetric matrix. Thus, we can reduce the number of unknown parameters in~\eqref{eq:min_admit} by following the approach of~\cite{ardakanian2017identification}. This is done by taking the lower triangular part of $\YY_\text{H}$ and then transforming it into a vector.

First, define the mapping $f: \Co^{u \times u} \mapsto \Co^{u(u+1)/2}$ which takes the lower triangular element of its input matrix and returns it as a vector. Second, define $\QQ_{\YY_\text{H}} \in \{0,1\}^{u^2 \times u(u+1)/2}$ such that $\ve \left( \YY_\text{H} \right) = \QQ_{\YY_\text{H}} f(\YY_\text{H})$~\cite{ardakanian2017identification}. The estimation problem can be re-written as,
\begin{equation*}
\min_{\xx \in \RR^{u(u+1)/2}} \left\|\ve \left( \II_\text{network}\right) - \left( \VV_\text{network} \Tr  \otimes \Id_u\right) \QQ_{\YY_\text{H}}\xx  \right\|^2, 
\label{eq:min_admit_vec}
\end{equation*}
where $\ve \left( \cdot \right)$ is the vectorization operator, $\otimes$ is the Kronecker product and $\Id_u \in \RR^{u \times u}$ is the identity matrix.

In~\cite{ardakanian2017identification}, the goal is also to identify the topology of the network. Here we instead estimate the harmonic admittances assuming full knowledge of the topology. Let $s=3(K+1)(N+ \frac{1}{2}\card \mathcal{M})$ be the number of unknown network parameters. Define a second mapping $g : \Co^{u\times u} \mapsto \Co^{s}$ which, similar to $f$, takes the lower triangular matrix of its input and returns it as a vector without the zero entries corresponding to pairs $(i,j) \notin \mathcal{M}$. 

Let $\mathbf{T} \in \{0,1\}^{u^2 \times s}$ be defined such that $f(\YY_\text{H}) = \mathbf{T} g(\YY_\text{H})$. The information about the topology of the network is preserved by using $\mathbf{T}$ as it encodes the location of the sparse entries of the harmonic admittance matrix. It thus permits to retrieve the original symmetric matrix from the output of $g$. The matrix $\mathbf{T}$ is obtained by inserting a row of zeros at row numbers corresponding to sparse entry of $f(\YY_\text{H})$ to the $u^2 \times u^2$ identity matrix. The original admittance matrix is then given by $f^{-1}\left(\mathbf{T} g\left(\YY_\text{H}\right)\right)$.

Using $\mathbf{T}$ and $g$, we constrain the entries of the harmonic admittance matrix corresponding to coordinates $(i,j) \notin \mathcal{M}$ to be zero, reducing the number of unknowns in the estimation problem. The final harmonic admittance estimation problem takes the following form:
\begin{equation*}
\min_{\yy \in \RR^{s}} \left\|\ve \left( \II_\text{network}\right) - \left( \VV_\text{network} \Tr  \otimes \Id_u\right) \QQ_{\YY_\text{H}} \mathbf{T} \yy  \right\|^2.
\label{eq:min_admit_final}
\end{equation*}
The estimated harmonic admittance parameters, $\yy_\text{est}$ , are given by
\begin{equation*}
\yy_\text{est} = \left( \mathbf{X}\Tr \mathbf{X} \right)^{-1} \mathbf{X} \Tr \ve \left( \II_\text{network} \right),
\end{equation*}
where $\mathbf{X} = \left( \VV_\text{network} \Tr  \otimes \Id_u\right) \QQ_{\YY_\text{H}} \mathbf{T}$. Lastly, the harmonic admittance matrix is given by $\YY_\text{H} = f^{-1} \left(\mathbf{T} \yy_\text{est} \right)$.

The problem has $s$ unknowns and $u$ equations. Each sample contains the information for the $3$ phases and $K+1$ frequencies. Thus, if $T \geq \frac{1}{2}\card \mathcal{M}$ then the problem is fully determined and the harmonic admittance matrix can be directly estimated in a noiseless setting.

\section{FCM estimation}
\label{sec:estimation}
We estimate the FCM using least squares. We consider a single converter, $n_\text{pc} \in \NN$, because the multiple converter case decouples into a collection of single converter problems. We assume that the node $n_\text{pc}$ is observable. We use harmonic current and voltage measurements to compute the estimate. Let $p = 6(K + 1)$ and $q = 6(K+1) + 1$, the dimensions of the measured vectors $\ii_\text{pc}$ and $\vv_\text{pc}$. The phasor $\ii_\text{pc}$ and the first $6(K+1)$ entries of $\vv_\text{pc}$ would typically be obtained from a PMU, and the last entry of $\vv_\text{pc}$, the dc current, from a smart meter.

In this section, we assume that the FCM does not vary over time. Given $\ii_{n_\text{pc},t}$ and $\vv_{n_\text{pc},t}$ for all $t=1,2, \ldots, T$, then provided that $T$ is sufficiently large, the frequency coupling matrix at node $n_\text{pc}$ can be estimated using the following convex program:
\begin{equation*}
\begin{aligned}
& \min_{\FF \in \RR^{p \times q}} & & \sum_{t=1}^T \left\|\ii_{n_\text{pc},t} - \FF \vv_{n_\text{pc},t} \right\|_2^2. \\
\end{aligned}
\end{equation*}
Equivalently, let $\II \in \RR^{p \times T}$ and $\VV \in \RR^{q \times T}$ be measurement matrices where column $t$ is the vector $\ii_{n_\text{pc},t}$ and $\vv_{n_\text{pc},t}$ respectively for $\II$ and $\VV$. Then the least squares problem is:
\begin{equation*}
\begin{aligned}
& \min_{\FF \in \RR^{p \times q}} & & \left\| \II - \FF \VV \right\|_F^2. \\
\end{aligned}
\label{eq:fcm:est}
\end{equation*}
We assume that the rows of $\VV$ are linearly independent. Given $T \geq q$, the problem is either fully determined ($T=q$) or over-determined ($T > q$). The FCM $\FF$ is then given by:
\begin{equation}
\FF = \II \VV \Tr \left(\VV \VV \Tr\right)^{-1} \label{eq:sol_ls}
\end{equation}
Note that if the rows of $\VV$ are linearly dependent, then the inverse is replaced by the pseudo-inverse in~\eqref{eq:sol_ls}.

Throughout this section, we assumed that we have access to $i_\text{dc}^0$. This assumption is mild because it can be estimated from metering or other measurements with lower resolution than PMUs.

\section{Online FCM estimation}

In this section, we give an online estimation algorithm for when the FCM varies through time. An example is a parking garage equipped with charging stations where electric cars are temporarily connected. Each newly connected or disconnected car would modify the garage's aggregate FCM. The FCM would therefore need to be continually updated to correctly describe the resulting harmonics. The online algorithm is shown in Algorithm~\ref{alg:iter_fcm}. At each time, we solve the following problem to estimate the FCM:
\[
\FF_{t} = \argmin_{\FF \in \RR^{p \times q}} \sum_{j=t+1}^{t+T} \left\|\ii_{n_\text{pc},j} - \FF \vv_{n_\text{pc},j} \right\|_2^2.
\]

The number of samples $T$ is fixed to some value greater than or equal to $q$, and the measurement matrices $\II_t$ and $\VV_t$ are iteratively updated to incorporate the $T$ most recent measurements (see Line 10 of Algorithm~\ref{alg:iter_fcm}). The Sherman-Morrison formula~\cite{sherman1950adjustment} is used to update the inverse matrix $\Vc_t=\left(\VV_t \VV_t \Tr \right)^{-1}$ in~\eqref{eq:sol_ls} using only algebraic operations~\cite{hager1989updating}. First, the oldest data from round $t-T$ are factored out of the inverse matrix on Line 7 in Algorithm~\ref{alg:iter_fcm}. Second, the new data collected at time $t$ are factored into the inverse matrix on Line 9. The costly inverse operation of~\eqref{eq:sol_ls} is only performed once during the initialization step. Finally, using the updated $\Vc_t$ matrix at time $t$ the estimate is obtained.

\begin{algorithm}
\begin{algorithmic}[1]
\STATE \textbf{Initialization:} Set $\II_0 \in \RR^{p \times T}$ and $\VV_0 \in \RR^{q \times T}$ using preliminary measurements. 
\STATE Compute $\Vc_0=\left( \VV_0 \VV_0 \Tr \right)^{-1}$.
\STATE Compute $\FF_0$ by solving~\eqref{eq:sol_ls}.
\medskip

\FOR{$t = 1,2, \ldots$}
\STATE Obtain $\ii_{n_\text{pc},t}$ and $\vv_{n_\text{pc},t}$.
\medskip

\STATEx \textbf{Factor out old measurements:}
\STATE Set $\cc = \VV_t(:,0)$ 
\STATE \[
\widetilde{\Vc}_t = \Vc_{t-1} + \frac{\Vc_{t-1}\cc \cc \Tr \Vc_{t-1}}{1 - \cc \Tr \Vc_{t-1} \cc}
\]
\STATEx \textbf{Factor in new measurements:}
\STATE Set $\dd = \vv_{n_\text{pc},t}$
\STATE \[
\Vc_t = \widetilde{\Vc}_t - \frac{\widetilde{\Vc}_t \dd \dd \Tr \widetilde{\Vc}_t}{1+ \dd \Tr \widetilde{\Vc}_t \dd}
\]
\STATE Update measurement matrices:
\begin{align*}
\II_t &= \begin{pmatrix}
\II_{t-1}\;(\text{ all },2 \text{ to last}) & \ii_{n_\text{pc},t}\;
\end{pmatrix} \\
\VV_t &= \begin{pmatrix}
\VV_{t-1}(\text{ all },2 \text{ to last}) & \vv_{n_\text{pc},t}
\end{pmatrix} 
\end{align*}
\STATE Update FCM:
\[
\FF_{t+1} = \II_t \VV_t \Vc_t 
\]
\ENDFOR
\end{algorithmic}
\caption{Online FCM estimation algorithm}
\label{alg:iter_fcm}
\end{algorithm}

\section{Numerical examples}

We now test each estimation problems and Theorem~\ref{thm:net_reduc} on the three node system shown in Figure~\ref{fig:network}. The system's parameters are given in Table~\ref{tab:para}. In all cases, the maximum harmonic order is $K=50$.	

\subsection{Line admittance estimation}
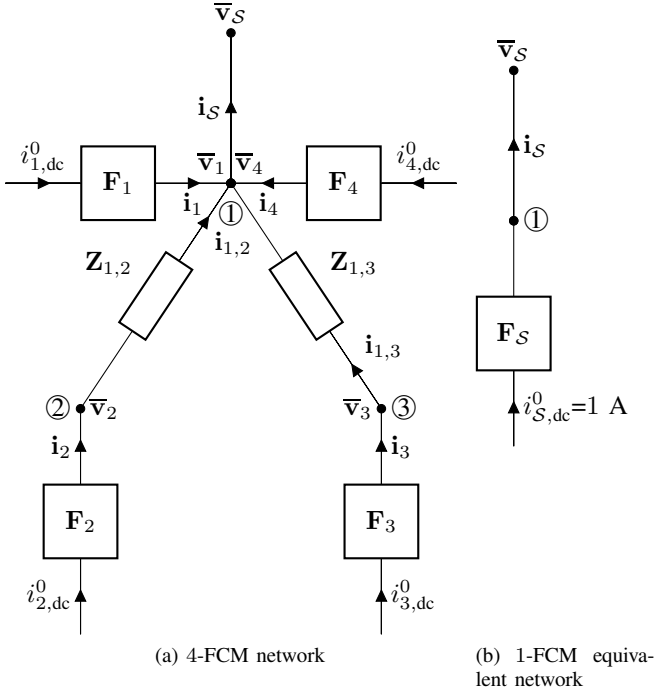
\begin{figure}[!t]
\centering
\subfloat[$4$-FCM network]{\centering
\begin{circuitikz}[american voltages]
	\ctikzset{label/align = straight}
	\draw
	  (0,0) to [twoport,t=$\FF_4$,*-, i<_=$\ii_4$] (3,0) 
	  (-3,0) to [twoport,t=$\FF_1$,-*, i>_=$i^\mathrm{dc}_1$,i_>=$\ii_1$] (0,0)
	  (0,0) to [short,i=$\ii_\SS$,*-*] (0,2) node[anchor=south]{$\ov_\SS$}
	  (-2,-3) to [generic=$\ZZ_{1,2}$,*-*,i_>=$\ii_{1,2}$] (0,0)
	  (0,0) to [generic=$\ZZ_{1,3}$,*-*,i^<=$\ii_{1,3}$] (2,-3)
	  (-2,-3) to [twoport,t=$\FF_2$, i<_=$\ii_2$] (-2,-6)
	  (2,-3) to [twoport,t=$\FF_3$,i<^=$\ii_3$] (2,-6)
	  (2,-6) to [short, i_>=$i^0_{3,\text{dc}}$] (2,-5)
	  (-2,-6) to [short, i^>=$i^0_{2,\text{dc}}$] (-2,-5)
	  (3,0) to [short, i_>=$i^0_{4,\text{dc}}$] (2,0)
	  (-3,0) to [short, i^>=$i^0_{1,\text{dc}}$] (-2,0)
	  (0,-0.15) node[anchor=north]{\raisebox{.5pt}{\textcircled{\raisebox{-.9pt} {1}}}}
	  (-2,-3) node[anchor=east]{\raisebox{.5pt}{\textcircled{\raisebox{-.9pt} {2}}}}
	  (2,-3) node[anchor=west]{\raisebox{.5pt}{\textcircled{\raisebox{-.9pt} {3}}}}
	  (2.,-3) node[anchor=east]{$\ov_3$}
	  (-2,-3) node[anchor=west]{$\ov_2$}
	  (-0.25,0) node[anchor=south]{$\ov_1$}
	  (0.25,0) node[anchor=south]{$\ov_4$};
\end{circuitikz}
\label{fig:network}}
\subfloat[$1$-FCM equivalent network]{\raisebox{25mm}{
\centering
\begin{circuitikz}[american voltages]
	\ctikzset{label/align = straight}
	\draw
	  (0,-3) to [twoport,t=$\FF_\SS$,-*, i>_=$i^0_{\SS,\text{dc}} \text{=} 1 \text{ A}$] (0,0)
	  (0,0) to [short,i_=$\ii_\SS$,*-*] (0,2) node[anchor=south]{$\ov_\SS$}
	  (0,0) node[anchor=west]{\raisebox{.5pt}{\textcircled{\raisebox{-.9pt} {1}}}}
	  ;

\end{circuitikz}
}
\label{fig:red}}
\caption{Example of the reduction theorem on a $4$-FCM network}
\label{fig:example_fcm}
\end{figure}

We first estimate the line admittances of the example in Figure~\ref{fig:network}. We generate the exact harmonic admittances using the resistances and susceptances given in Table~\ref{tab:para} for the fundamental frequency. We assume that the lines are purely inductive. We set the impedance of each line at harmonic frequencies to $\mathbf{z}_{m,n}(p) = r_{m,n}(p) + j k  x_{m,n}(p)$ for all $(i,j) \in \mathcal{M}$, $k=0,2,3,\ldots,K$ and $p=a,b,c$. The impedance is then used to compute the admittance. Note that a more accurate model, e.g., one based on the steady state solution to the telegrapher's equations or accounting for the skin effect, could equivalently be used to set the harmonic impedances.

The node voltages are sampled from a normal distribution at each time $t$. The mean voltage for $k=1$ is provided in Table~\ref{tab:voltages}. For $0^{\text{th}}$ harmonic, only the real part of the mean is used. For $k \neq 0$, this mean is divided by $1.1^k$ to obtain different values across the frequencies. The standard variation of the normal distribution for frequency $k$ is $0.005/1.1^k$.

\begin{table}[tb]
\renewcommand{\arraystretch}{1.3}
  \caption{3-node network parameters}
  \label{tab:para}
  \centering

  \begin{tabular}{ccc}
  \hline

  \hline
  \textbf{Parameter} & \textbf{Value}  & \textbf{Unit} \\
  \hline
  $i^0_{1,\text{dc}}$ & 0.05 & A \\
  $i^0_{2,\text{dc}}$ & 0.025 & A \\
  $i^0_{3,\text{dc}}$ & 0.075 & A \\
  $i^0_{4,\text{dc}}$ & 0.06 & A \\
  $r_{1,2}(a)$ & 0.05 & $\Omega$ \\
  $r_{1,2}(b)$ & 0.06 & $\Omega$ \\
  $r_{1,2}(c)$ & 0.04 & $\Omega$ \\
  $r_{1,3}(a)$ & 0.075 & $\Omega$ \\
  $r_{1,3}(b)$ & 0.08 & $\Omega$ \\
  $r_{1,3}(c)$ & 0.07 & $\Omega$ \\
  $x_{1,2}(a)$ & 0.1 & $\Omega$ \\
  $x_{1,2}(b)$ & 0.95 & $\Omega$\\
  $x_{1,2}(c)$ & 0.15 & $\Omega$\\
  $x_{1,3}(a)$ & 0.15 & $\Omega$\\
  $x_{1,3}(b)$ & 0.145 & $\Omega$\\
  $x_{1,3}(c)$ & 0.155 & $\Omega$ \\
  \hline
  \hline
  \end{tabular}
\end{table}

\begin{table}[tb]
\renewcommand{\arraystretch}{1.3}
  \caption{Mean voltages at fundamental frequency for harmonic admittance estimation}
  \label{tab:voltages}
  \centering
  \begin{tabular}{c c c c}
  \hline
  \hline
  \textbf{Mean voltage}              & \textbf{Phase $a$ [V]} & \textbf{Phase $b$ [V]} & \textbf{Phase $c$ [V]} \\
  \hline
  $\overline{v}_1^1$  & $1.25 + 0.625j$ & $1 + 0.5j$ & $0.75 + 0.375j$  \\
  $\overline{v}_2^1$  & $2.5 + 0.125j$ & $2 + 0.1j$ & $1.5 + 0.075j$  \\
  $\overline{v}_3^1$  & $0.625 + 1.25$ & $0.5 + j$ & $0.375 + 0.75j$ \\
  \hline

  \hline
  \end{tabular}
\end{table}

The harmonic current is set to $\ii_{\text{bus},t} = \YY_\text{H}^\ast \vv_{\text{bus},t}$ where $\YY_\text{H}^\ast$ is the exact harmonic admittance matrix. Zero-mean Gaussian noise is then added to the harmonic current and voltage to model measurement errors. The standard deviation of the noise is set to be a percentage of the mean voltage or current. We vary this percentage in the simulation. 

\begin{figure}[!t]
\centering
\subfloat[Relative estimation error as a function of the noise variance for $T=10$]{\includegraphics[width=1\columnwidth]{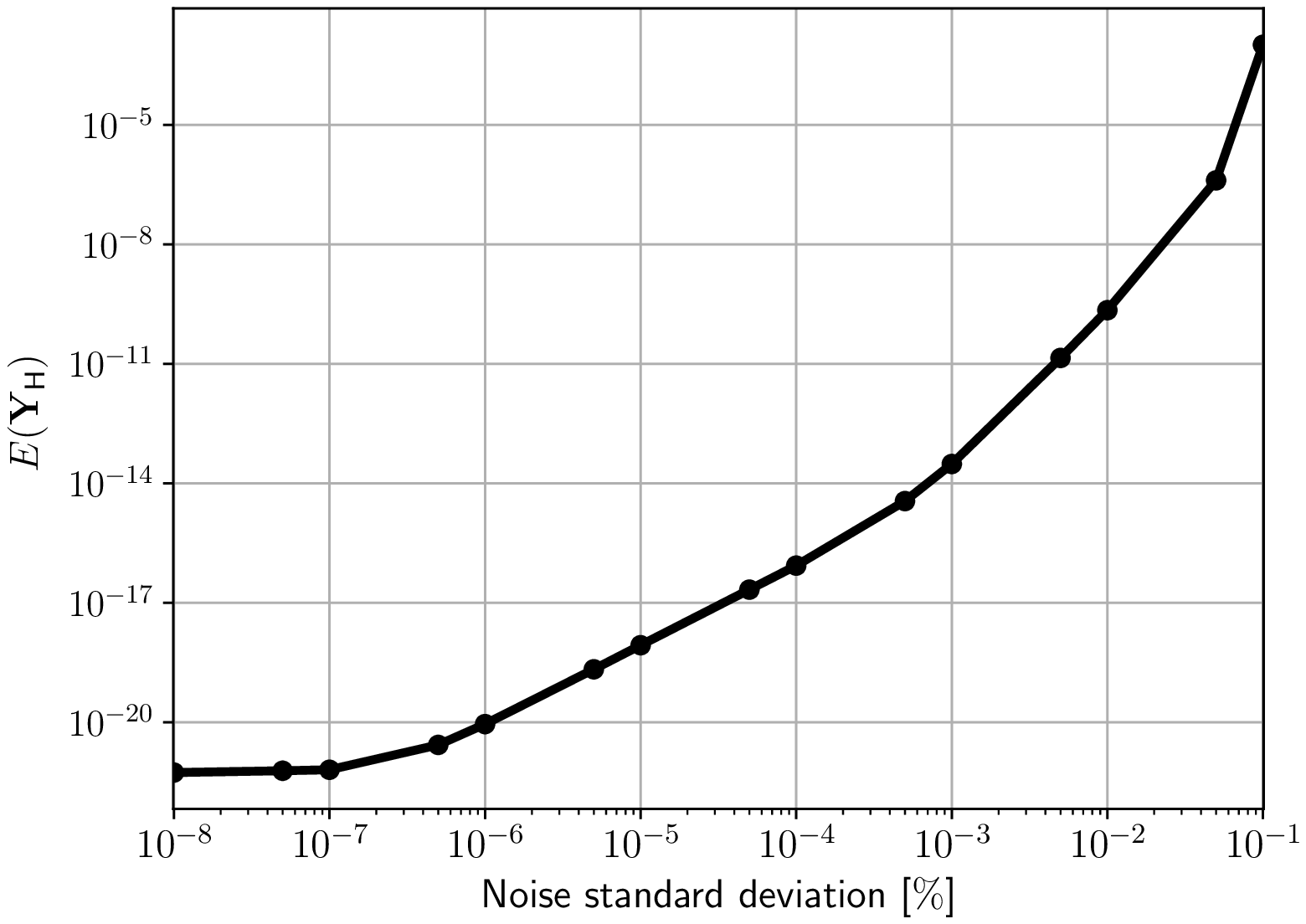}%
\label{fig:e_noise_var}}
\vspace{-0.4cm}

\subfloat[Relative estimation error as a function of $T$ under $1\%$ noise]{\includegraphics[width=1\columnwidth]{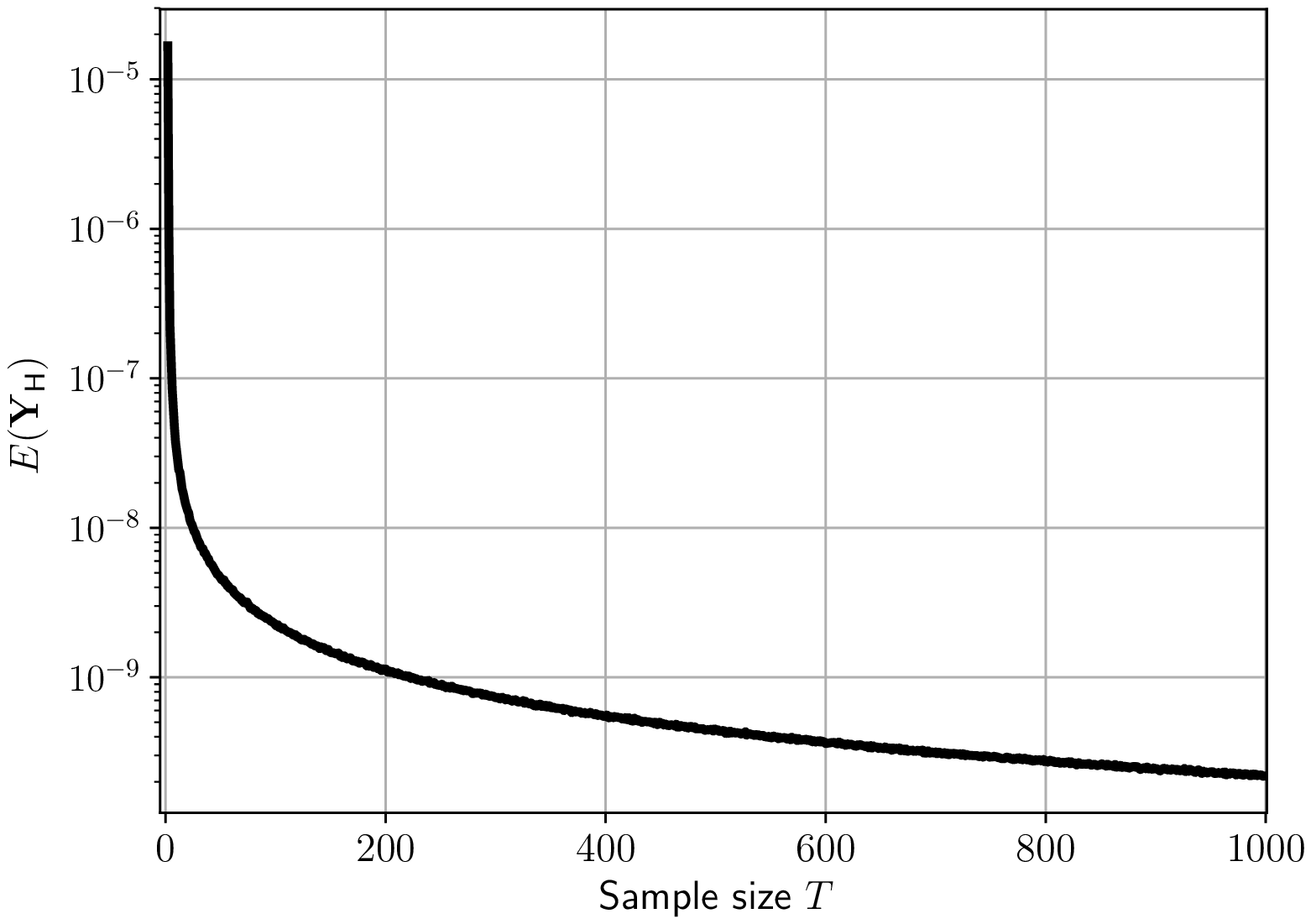}%
\label{fig:e_t}}
\caption{Harmonic admittance relative estimation error averaged over 100 simulations}
\label{fig:Y_Error}
\end{figure}

The relative error of an estimate is defined as:
\begin{equation*}
E(\YY_\text{H}) = \frac{\| \YY_\text{H}^\ast - \YY_\text{H} \|^{2}_F}{\| \YY_\text{H}^\ast \|^{2}_F}.
\end{equation*}
We present the relative estimation error of the harmonic line admittance matrix averaged over of $100$ simulations in Figure~\ref{fig:Y_Error}. In Figure~\ref{fig:e_noise_var}, the estimation error is shown as a function of the standard deviation of the noise for $T=10$. As anticipated, the error increases with the standard deviation of the noise. To improve the performance under high variance noise, $T$ can be increased. Figure~\ref{fig:e_t} shows the relative estimation error as a function of the sample size, $T$, with $1\%$ noise standard deviation. As expected, increasing the length of the sampling window can decrease error.

\subsection{Batch FCM estimation}
\label{sub:setting}

\begin{figure}[!t]
  \centering
  \centering
\begin{circuitikz}[american voltages]
	\ctikzset{label/align = straight}
	\draw
	  (-4,0) to [twoport,t=$\FF_e$,-*, i_>=$\ii_{e}$] (0,0)
	  (0,0) node[anchor=south]{\raisebox{.5pt}{\textcircled{\raisebox{-.9pt} {$e$}}}} node[anchor=north]{$\ov_e$}
	  (-4,0) to [short, i>_=$i^0_{e,\text{dc}}$] (-2.5,0)
	  ;

\end{circuitikz}
  \caption{Single power converter FCM estimation}
  \label{fig:est}
\end{figure}
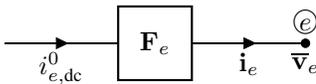

We now estimate the FCM of the power converter shown in Figure~\ref{fig:est}. In our numerical simulations, we sample the input harmonic voltages at each time from a normal distribution with mean $\overline{\vv}^k_e$ and standard deviation $0.005$ for the real and imaginary components for each harmonic $k$. The mean harmonic voltage $\overline{\vv}^k_e$ is set to the input voltages used in~\cite[Table II]{lehn2007frequency}.

We solve for the harmonic currents using $\ii_{e,t} = \FF^\ast_{e} \vv_{n_\text{pc},t}$, where $\FF^\ast_{e}$ denotes the exact FCM computed using the calculation described in~\cite{lehn2007frequency}. 
To simulate measurement errors, zero-mean Gaussian noise with variances corresponding to $0.1\%$ or $1\%$ of the mean of the harmonic current and voltage norm are added to each component of $\ii_{e,t}$ and $\vv_{e,t}$.

To compute the FCM, internal component values (resistance, inductance of each phase, capacitance), the switching times and sequence are required. These are set according to~\cite[Table I]{lehn2007frequency}. The switching times are sampled uniformly between 0 and $2\pi$ and each element of a switching sequence is sampled according to a Bernoulli distribution with probability one half. Each time a new FCM is needed, the switching times and sequence are re-sampled while the internal parameters are kept constant.

For an estimated FCM $\FF_e$, the relative estimation error $E_e$ is given by:
\begin{equation}
E(\FF_e) = \frac{\left\| \FF^\ast_e - \FF_e \right\|_F^2}{\left\|\FF^\ast_e \right\|_F^2}.
\label{eq:estimation_error}
\end{equation}

\begin{figure}[!t]
\centering
\subfloat[$0.1\%$ measurement noise]{\includegraphics[width=1\columnwidth]{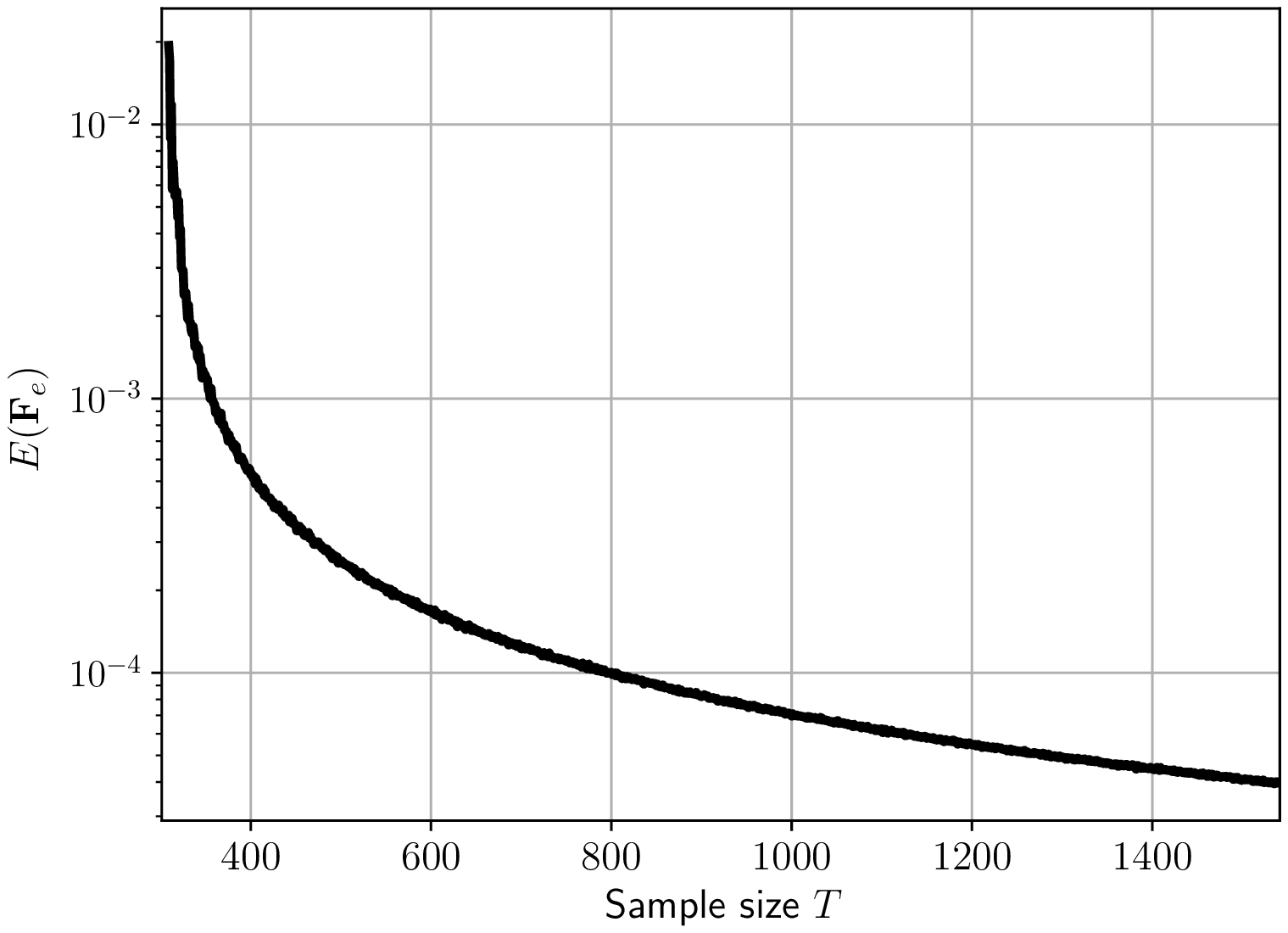}%
\label{fig:wo}}
\vspace{-0.4cm}

\subfloat[$1\%$ measurement noise]{\includegraphics[width=1\columnwidth]{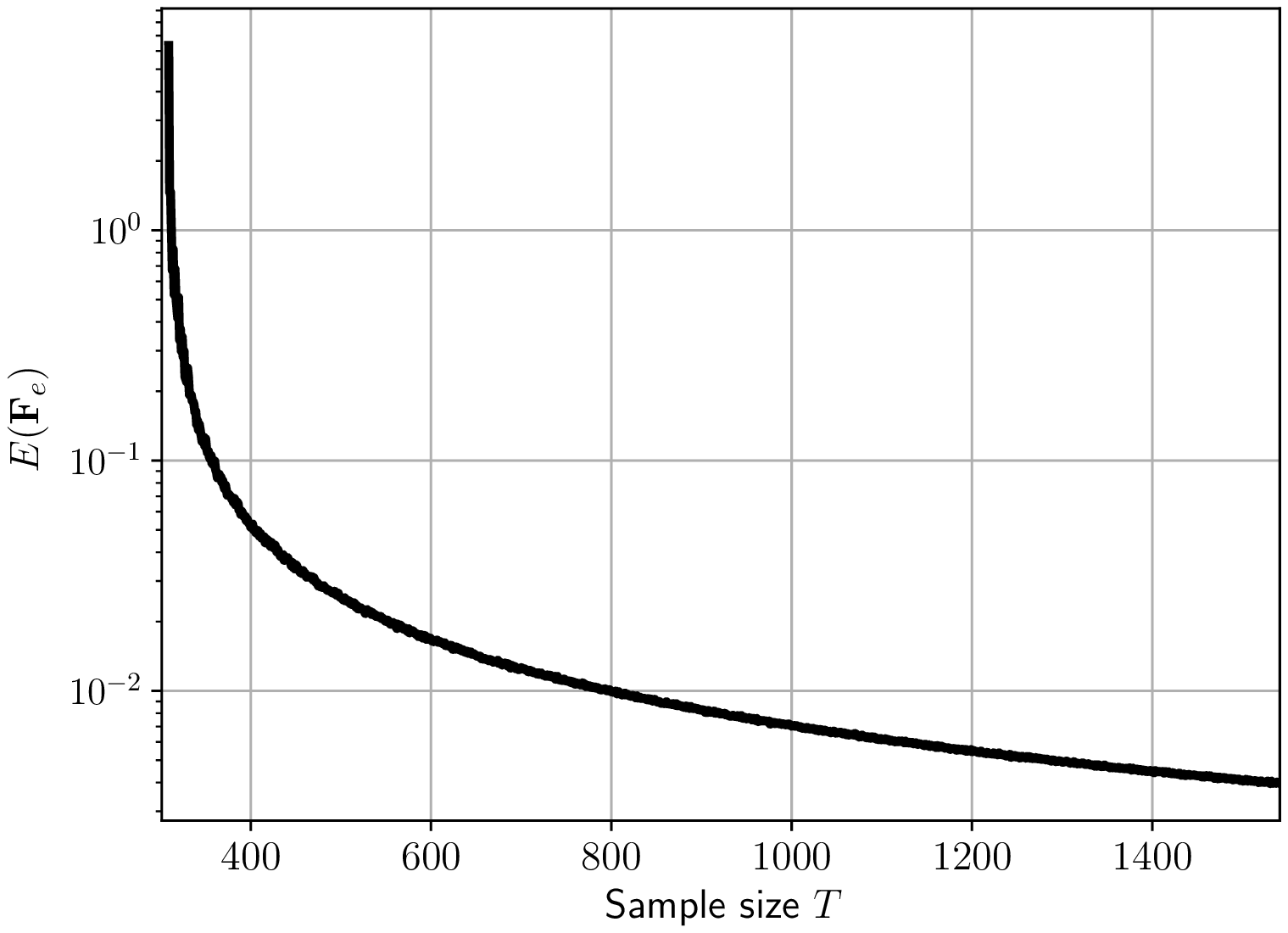}%
\label{fig:wo_small}}
\caption{Batch estimation relative error for $T= q+1$ to $5q$}
\label{fig:inst_est}
\end{figure}

The operator desires an instantaneous estimate of the converter's FCM. Note that by invoking Theorem~\ref{thm:net_reduc} and Corollary~\ref{cor:tree_reduc}, the FCM at the node could represent the aggregate FCM for a downstream subtree. The operator measures harmonic voltages, currents, and dc current for $T$ rounds before the estimate is needed. The dc current, $i_{e,\text{dc}}^0$, is set to $5$~mA for the numerical simulations and is subject to the same noise as the input voltage. We present performance results without measurement noise, $0.1\%$ and $1\%$ measurement noise. The estimation error~\eqref{eq:estimation_error} is given in Figure~\ref{fig:inst_est} as a function of $T$ for the $0.1\%$ and $1\%$ noise level of noise. Without noise, the estimation error is virtually zero. We see that the error decreases with the noise and as $T$ increases.

Figure~\ref{fig:inst_est} shows error when there is observation noise. The estimation error is below $0.01\%$ of the Frobenius norm of the exact FCM when $T$ is greater than $2.3q$. Under higher noise, a larger value of $T$ can be used to obtain similar performance. For example, when we set $T=165q$ and run the simulation $100$ times under $1\%$ noise, the average estimation error is $9.69 \times 10^{-5}$. This shows that similar performance can be obtained when subject to strong measurement noise if the number of samples $T$ is large enough.

\subsection{Online FCM estimation}
\label{sub:real-time_example}

\begin{figure}[h]
  \centering
  \includegraphics[width=1\columnwidth]{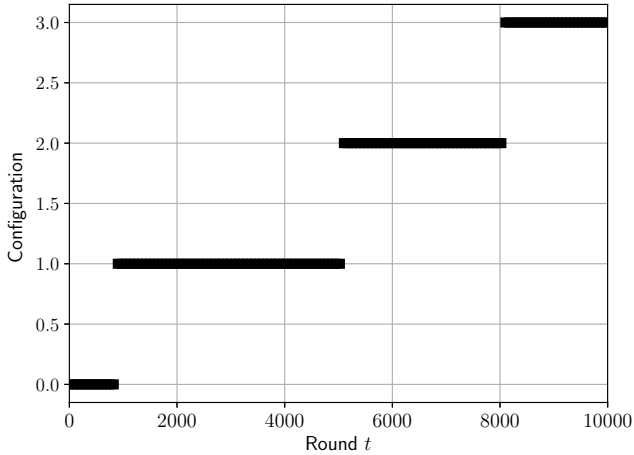}
  \caption{Power converter configuration as a function of $t$}
  \label{fig:changes}
\end{figure}

We now assume that the FCM changes with time. We replace $\FF_e$ with $\FF_{e,t}$ at time $t$ in Figure~\ref{fig:est}. We consider a $0.1\%$ observation noise and horizon $T=2q$. As previously mentioned, this could represent a scenario where electric vehicles can come and go at a parking garage. We set the time horizon to $10^4$ and consider $4$ different configurations during this time horizon as shown in Figure~\ref{fig:changes}. Each of the four FCM configurations corresponds to a different duty cycle, as described in the previous section. The relative estimation error for the online case is given by
\begin{equation}
E_t(\FF_{e,t}) = \frac{\left\| \FF^\ast_{e,t} - \FF_{e,t} \right\|_F^2}{\max_\tau \left\|\FF^\ast_{e,\tau}\right\|_F^2}.
\end{equation}

\begin{figure}[h]
  \centering
  \includegraphics[width=1\columnwidth]{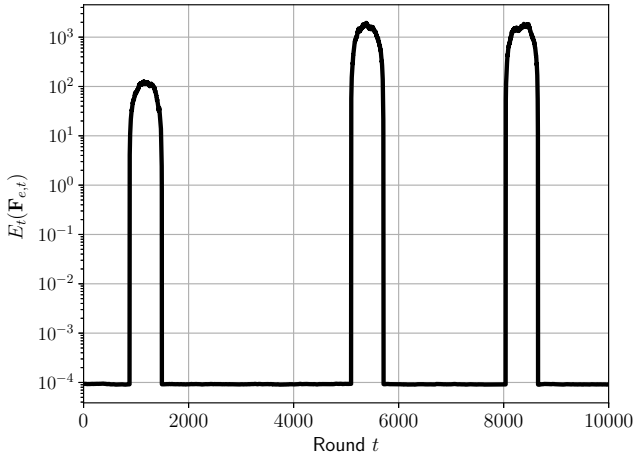}
  \caption{Online FCM estimation relative estimation error for $T=2q$}
  \label{fig:it_s_est}
\end{figure}

The results for the relative estimation error are presented in Figure~\ref{fig:it_s_est}. The estimator performs well except during brief transitions between configurations. This is because the measurement matrices have data from the previous configuration. To remedy this, the estimator could be combined with an event detection algorithm to omit prior measurements when a significant change is occurring.


\subsection{Network reduction example}
In this section, we apply the reduction theorem to the network Figure~\ref{fig:network} and obtain the equivalent, virtual FCM $\FF_\SS$ of Figure~\ref{fig:red}. Two FCMs, $\FF_1$ and $\FF_2$ are directly connected to node $1$. Two other FCMs, $\FF_2$ at node $2$ and $\FF_3$ at node $3$, are connected to node $1$ via lines with impedance $\ZZ_{1,2}$ and $\ZZ_{1,3}$ respectively.

We set the average value of the dc currents, resistances, susceptances according to Table~\ref{tab:para} and $\ov_\SS$ according to~\cite[Table II]{lehn2007frequency}. To the each dc current, resistances, susceptances and components of $\ov_\SS$, we add a zero-mean Gaussian noise with standard deviation given respectively by $0.005$, $0.01$, $0.01$, $0.005$. The four exact FCMs have the same internal parameters and randomly sampled switching sequences as in Section~\ref{sub:setting}. We omit measurement noise as our objective is to validate the theoretical results of Section~\ref{sec:network_red}. 

We run $250$ tests. In each test, we compute the equivalent reduced FCM, $\FF_\SS$, using Theorem~\ref{thm:net_reduc} given all four FCMs and their dc input current. We estimate the equivalent reduced FCM, $\FF_\SS^\text{estimation}$, using Section~\ref{sec:estimation} with $T=2q$. The exact harmonic current $\ii_\SS$ is calculated using the network equations given the four FCMs, their dc currents and $\ov_\SS$. The details for the computation of $\ii_\SS$ are given in Appendix~\ref{app:network_eq}. We also compute the harmonic current obtained with the equivalent reduced FCM, $\ii_\SS^\text{reduction} = \FF_\SS \vv_\SS$, and that obtained from the estimated FCM, $\ii_\SS^\text{estimated} = \FF^\text{estimated}_\SS \vv_\SS$. Table~\ref{tab:perf} gives the mean FCM error as well as the following current error metrics:
\begin{align*}
\varepsilon^\text{reduction} &= \frac{\left\| \ii_\SS - \ii_\SS^\text{reduction} \right\|}{\left\|\ii_\SS\right\|},\\
\varepsilon^\text{estimated} &= \frac{\left\| \ii_\SS - \ii_\SS^\text{estimated} \right\|}{\left\|\ii_\SS\right\|}, \\
\varepsilon^\text{comparison} &= \frac{\left\| \ii_\SS^\text{reduction} - \ii_\SS^\text{estimated} \right\|}{\left\|\ii_\SS^\text{estimated}\right\|}.
\end{align*}
All errors are effectively zero, in accordance with the theoretical results of Section~\ref{sec:network_red}.

\begin{table}[tb]
\renewcommand{\arraystretch}{1.3}
  \caption{Validation performance for $250$ tests}
  \label{tab:perf}
  \centering
  \begin{tabular}{c c}
  \hline
  \hline
  \textbf{Mean error}              & \textbf{Value}  \\
  \hline
  $\overline{\varepsilon}^\text{reduction}$  &  $1.23 \times 10^{-15}$  \\
  $\overline{\varepsilon}^\text{estimated}$  &  $1.99 \times 10^{-11}$ \\
  $\overline{\varepsilon}^\text{comparison}$   &  $1.99 \times 10^{-11}$ \\
  $\overline{E}(\FF^\text{estimated}_\SS)$       &  $9.28\times 10^{-23}$ \\
  \hline

  \hline
  \end{tabular}
\end{table}

\section{Conclusion}
We have presented basic approaches for estimating network parameters necessary for modeling harmonics. We have posed least squares problems for estimating the line admittances and FCMs, and an online algorithm for the latter case. We have also given a network reduction theorem, which enables one to model arbitrary, unobservable subtrees with a single, virtual FCM. We have validated all methods on a simple numerical example.

\appendices

\section{Network equation solution}
\label{app:network_eq}
In this appendix, we solve for the relevant electrical quantities in the $4$-FCM network in Figure~\ref{fig:network}. All voltages are the same at a given node, and hence $\ov_\SS = \ov_1 = \ov_4$. By Kirchhoff's current law, we have $\ii_{1,2} = \ii_2$ and $\ii_{1,3} = \ii_3$. There are 8 unknown vectors: $\ii_1$, $\ii_2$, $\ii_3$, $\ii_4$, $\ov_\SS$, $\ov_2$, $\ov_3$. We set $\vv_\SS$ as our input data. 

For the 4-FCM network, we have the following equations. By Kirchhoff's current law at node $1$, we have
\begin{equation}
\ii_1 + \ii_2 + \ii_3 + \ii_4 = \ii_\SS. \label{eq:network_first}
\end{equation}
Using Ohm's law on line $(1,2)$, we obtain:
\begin{equation}
\vv_2 - \vv_1 = \ZZ_{1,2} \ii_2,
\end{equation}
and for line $(1,3)$,
\begin{equation}
\vv_3 - \vv_1 = \ZZ_{1,3} \ii_3.
\end{equation}
We apply the FCM relation at the three nodes:
\begin{align}
\ii_1 &= \FF_1 \begin{pmatrix}
\ov_1 \\ i^\mathrm{dc}_1
\end{pmatrix} = \oF \ov_\SS + \ff_1 i_{1,\text{dc}}^0, \\
\ii_2 &= \FF_2 \begin{pmatrix}
\ov_2 \\ i^\mathrm{dc}_2
\end{pmatrix} = \oF \ov_2 + \ff_2 i_{2,\text{dc}}^0, \\
\ii_3 &= \FF_3 \begin{pmatrix}
\ov_3 \\ i^\mathrm{dc}_3
\end{pmatrix} = \oF \ov_3 + \ff_3 i_{3,\text{dc}}^0, \\
\ii_4 &= \FF_4 \begin{pmatrix}
\ov_4 \\ i^\mathrm{dc}_4
\end{pmatrix} = \oF \ov_\SS + \ff_4 i_{4,\text{dc}}^0. \label{eq:network_last}
\end{align}

Fixing $\vv_\SS$ at a measured value, we have 7 unknown vectors and 7 sets of equations. Let $\Id_p \in \RR^{p \times p}$ be the identity matrix, $\uZ \in \RR^{p \times p}$ a matrix made only of zeros and $\ze \in \RR^p$ a vector made only of zeros. We rewrite~\eqref{eq:network_first}--\eqref{eq:network_last} as 
\begin{equation}
\begin{aligned}
\begin{pmatrix}
\Id_p & \Id_p & \Id_p & \Id_p & -\Id_p & \uZ & \uZ \\
\uZ & \ZZ_{1,2} & \uZ & \uZ & \uZ & \Id_p & \uZ \\
\uZ & \uZ & \ZZ_{1,3} & \uZ & \uZ & \uZ & \Id_p \\
\Id_p & \uZ & \uZ & \uZ & \uZ & \uZ & \uZ \\
\uZ & \Id_p & \uZ & \uZ & \uZ & -\oF_2 & \uZ \\
\uZ & \uZ & \Id_p & \uZ & \uZ & \uZ & -\oF_3 \\
\uZ & \uZ & \uZ & \Id_p & \uZ & \uZ & \uZ 
\end{pmatrix}
\begin{pmatrix}
\ii_1 \\ \ii_2 \\ \ii_3 \\ \ii_4 \\ \ii_\SS \\ \ov_1 \\ \ov_2
\end{pmatrix}
=\\
\begin{pmatrix}
\ze \\ \ov_\SS \\ \ov_\SS \\ \oF_\SS \ov_1 + \ff_1 i_{1,\text{dc}}^0 \\ \ff_2 i_{2,\text{dc}}^0 \\ \ff_3 i_{3,\text{dc}}^0 \\ \oF_\SS \ov_\SS + \ff_4 i_{4,\text{dc}}^0
\end{pmatrix},
\end{aligned}
\label{eq:network_syst}
\end{equation}
and solve for the 7 unknown vectors. From the solution of~\eqref{eq:network_syst}, we obtain the exact value of $\ii_\SS$.



\begin{thebibliography}{10}
\providecommand{\url}[1]{#1}
\csname url@samestyle\endcsname
\providecommand{\newblock}{\relax}
\providecommand{\bibinfo}[2]{#2}
\providecommand{\BIBentrySTDinterwordspacing}{\spaceskip=0pt\relax}
\providecommand{\BIBentryALTinterwordstretchfactor}{4}
\providecommand{\BIBentryALTinterwordspacing}{\spaceskip=\fontdimen2\font plus
\BIBentryALTinterwordstretchfactor\fontdimen3\font minus
  \fontdimen4\font\relax}
\providecommand{\BIBforeignlanguage}[2]{{%
\expandafter\ifx\csname l@#1\endcsname\relax
\typeout{** WARNING: IEEEtran.bst: No hyphenation pattern has been}%
\typeout{** loaded for the language `#1'. Using the pattern for}%
\typeout{** the default language instead.}%
\else
\language=\csname l@#1\endcsname
\fi
#2}}
\providecommand{\BIBdecl}{\relax}
\BIBdecl

\bibitem{arrillaga2004power}
J.~Arrillaga and N.~R. Watson, \emph{Power system harmonics}.\hskip 1em plus
  0.5em minus 0.4em\relax John Wiley \& Sons, 2004.

\bibitem{carrasco2006power}
J.~M. Carrasco, L.~G. Franquelo, J.~T. Bialasiewicz, E.~Galv{\'a}n, R.~C.
  {P}ortillo {G}uisado, M.~M. Prats, J.~I. Le{\'o}n, and N.~Moreno-Alfonso,
  ``Power-electronic systems for the grid integration of renewable energy
  sources: A survey,'' \emph{IEEE {T}ransactions on {I}ndustrial
  {E}lectronics}, vol.~53, no.~4, pp. 1002--1016, 2006.

\bibitem{masters2013renewable}
G.~M. Masters, \emph{Renewable and efficient electric power systems}.\hskip 1em
  plus 0.5em minus 0.4em\relax John Wiley \& Sons, 2013.

\bibitem{de2006harmonics}
F.~De~La~Rosa, \emph{Harmonics and power systems}.\hskip 1em plus 0.5em minus
  0.4em\relax CRC Press Boca Raton, 2006.

\bibitem{henderson1994harmonics}
R.~D. Henderson and P.~J. Rose, ``Harmonics: The effects on power quality and
  transformers,'' \emph{IEEE {T}ransactions on {I}ndustry {A}pplications},
  vol.~30, no.~3, pp. 528--532, 1994.

\bibitem{gray2012time}
P.~Gray and P.~Lehn, ``Time-domain derived frequency-domain voltage source
  converter model for harmonic analysis,'' in \emph{Harmonics and Quality of
  Power (ICHQP), 15th International Conference on}.\hskip 1em plus 0.5em minus
  0.4em\relax IEEE, 2012, pp. 512--517.

\bibitem{lian2012steady}
K.~L. Lian and P.~Lehn, ``Steady-state simulation methods of closed-loop power
  converter systems—a systematic solution procedure,'' \emph{IEEE
  {T}ransactions on {C}ircuits and {S}ystems {I}: {R}egular {P}apers}, vol.~59,
  no.~6, pp. 1299--1311, 2012.

\bibitem{rajagopal1993harmonic}
N.~Rajagopal and J.~Quaicoe, ``Harmonic analysis of three-phase ac/dc
  converters using the harmonic admittance method,'' in \emph{Electrical and
  Computer Engineering, Canadian Conference on}.\hskip 1em plus 0.5em minus
  0.4em\relax IEEE, 1993, pp. 313--316.

\bibitem{fauri1997harmonic}
M.~Fauri, ``Harmonic modelling of non-linear load by means of crossed frequency
  admittance matrix,'' \emph{IEEE {T}ransactions on {P}ower {S}ystems},
  vol.~12, no.~4, pp. 1632--1638, 1997.

\bibitem{sun2007harmonically}
Y.~Sun, G.~Zhang, W.~Xu, and J.~G. Mayordomo, ``A harmonically coupled
  admittance matrix model for ac/dc converters,'' \emph{IEEE {T}ransactions on
  {P}ower {S}ystems}, vol.~22, no.~4, pp. 1574--1582, 2007.

\bibitem{lehn2007frequency}
P.~Lehn and K.~Lian, ``Frequency coupling matrix of a voltage-source converter
  derived from piecewise linear differential equations,'' \emph{IEEE
  {T}ransactions on {P}ower {D}elivery}, vol.~22, no.~3, pp. 1603--1612, 2007.

\bibitem{yahyaie2014application}
F.~Yahyaie, P.~Gray, and P.~Lehn, ``Application of experimentally measured
  frequency coupling matrices for improved harmonic estimation,'' in
  \emph{Proc. 9th Annual CIGR\'E Canada Conf.}, 2014, pp. 290--294.

\bibitem{zong2016new}
X.~J. Zong, P.~A. Gray, and P.~W. Lehn, ``New metric recommended for {IEEE}
  {S}tandard 1547 to limit harmonics injected into distorted grids,''
  \emph{IEEE {T}ransactions on {P}ower {D}elivery}, vol.~31, no.~3, pp.
  963--972, 2016.

\bibitem{yahyaie2016using}
F.~Yahyaie and P.~W. Lehn, ``Using frequency coupling matrix techniques for the
  analysis of harmonic interactions,'' \emph{IEEE {T}ransactions on {P}ower
  {D}elivery}, vol.~31, no.~1, pp. 112--121, 2016.

\bibitem{saadeh2016estimation}
M.~Saadeh, M.~Alsarray, and R.~McCann, ``Estimation of the bus admittance
  matrix for transmission systems from synchrophasor data,'' in
  \emph{Transmission and Distribution Conference and Exposition (T\&D), 2016
  IEEE/PES}, pp. 1--5.

\bibitem{tian2017harmonic}
Y.~Tian, J.~A. Taylor, and N.~Li, ``Harmonic reduction via optimal power flow
  and the frequency coupling matrix,'' in \emph{Control Technology and
  Applications (CCTA), 2017 IEEE Conference on}, pp. 2150--2157.

\bibitem{larsen1989low}
E.~Larsen, D.~Baker, and J.~McIver, ``Low-order harmonic interactions on ac/dc
  systems,'' \emph{IEEE {T}ransactions on {P}ower {D}elivery}, vol.~4, no.~1,
  pp. 493--501, 1989.

\bibitem{jalali1991harmonic}
S.~Jalali and R.~Lasseter, ``Harmonic interaction of power systems with static
  switching circuits,'' in \emph{Power Electronics Specialists Conference, 22nd
  Annual IEEE}, 1991, pp. 330--337.

\bibitem{saniter2003modelling}
C.~Saniter, A.~Wood, R.~Hanitsch, and D.~Schulz, ``Modelling the effects of
  {AC} system impedance unbalance on {PWM} converters using frequency coupling
  matrices,'' in \emph{Power Tech Conference Proceedings, IEEE Bologna},
  vol.~2, 2003, pp. 1--6.

\bibitem{hu1992harmonic}
L.~Hu and R.~Yacamini, ``Harmonic transfer through converters and hvdc links,''
  \emph{IEEE {T}ransactions on {P}ower {E}lectronics}, vol.~7, no.~3, pp.
  514--525, 1992.

\bibitem{bazrafshan2018comprehensive}
M.~Bazrafshan and N.~Gatsis, ``Comprehensive modeling of three-phase
  distribution systems via the bus admittance matrix,'' \emph{IEEE
  {T}ransactions on {P}ower {S}ystems}, vol.~33, no.~2, pp. 2015--2029, 2018.

\bibitem{ardakanian2017identification}
O.~Ardakanian, V.~W. Wong, R.~Dobbe, S.~H. Low, A.~von Meier, C.~Tomlin, and
  Y.~Yuan, ``On identification of distribution grids,'' \emph{arXiv preprint
  arXiv:1711.01526}, 2017.

\bibitem{carta2009pmu}
A.~Carta, N.~Locci, and C.~Muscas, ``A {PMU} for the measurement of
  synchronized harmonic phasors in three-phase distribution networks,''
  \emph{IEEE {T}ransactions on {I}nstrumentation and {M}easurement}, vol.~58,
  no.~10, pp. 3723--3730, 2009.

\bibitem{jain2017fast}
S.~K. Jain, P.~Jain, and S.~N. Singh, ``A fast harmonic phasor measurement
  method for smart grid applications,'' \emph{IEEE {T}ransactions on {S}mart
  {G}rid}, vol.~8, no.~1, pp. 493--502, 2017.

\bibitem{melo2017harmonic}
I.~D. Melo, J.~L. Pereira, A.~M. Variz, and P.~A. Garcia, ``Harmonic state
  estimation for distribution networks using phasor measurement units,''
  \emph{Electric {P}ower {S}ystems {R}esearch}, vol. 147, pp. 133--144, 2017.

\bibitem{phadke2008synchronized}
A.~G. Phadke and J.~S. Thorp, \emph{Synchronized phasor measurements and their
  applications}.\hskip 1em plus 0.5em minus 0.4em\relax Springer, 2008.

\bibitem{martin2008exploring}
K.~Martin, D.~Hamai, M.~Adamiak, S.~Anderson, M.~Begovic, G.~Benmouyal,
  G.~Brunello, J.~Burger, J.~Cai, B.~Dickerson \emph{et~al.}, ``Exploring the
  {IEEE} {s}tandard {C}37. 118--2005 synchrophasors for power systems,''
  \emph{IEEE {T}ransactions on {P}ower {D}elivery}, vol.~23, no.~4, pp.
  1805--1811, 2008.

\bibitem{abur2004power}
A.~Abur and A.~G. Exposito, \emph{Power system state estimation: theory and
  implementation}.\hskip 1em plus 0.5em minus 0.4em\relax CRC Press, 2004.

\bibitem{krumpholz1980power}
G.~Krumpholz, K.~Clements, and P.~Davis, ``Power system observability: a
  practical algorithm using network topology,'' \emph{IEEE Transactions on
  {P}ower {A}pparatus and {S}ystems}, no.~4, pp. 1534--1542, 1980.

\bibitem{sherman1950adjustment}
J.~Sherman and W.~J. Morrison, ``Adjustment of an inverse matrix corresponding
  to a change in one element of a given matrix,'' \emph{The Annals of
  Mathematical Statistics}, vol.~21, no.~1, pp. 124--127, 1950.

\bibitem{hager1989updating}
W.~W. Hager, ``Updating the inverse of a matrix,'' \emph{SIAM {R}eview},
  vol.~31, no.~2, pp. 221--239, 1989.

\end{thebibliography}
\end{document}